# Hyperbranched Polyesters by Polycondensation

# of Fatty Acid-based AB$_n$-type Monomers

Blandine Testud,[a,b] Didier Pintori,[c] Etienne Grau,[a,b] Daniel Taton[a,b*] and Henri Cramail[a,b*]

a. Université de Bordeaux, Laboratoire de Chimie des Polymères Organiques, UMR 5629, Bordeaux INP/ENSCBP, 16 avenue Pey-Berland, F-33607 Pessac Cedex, France

b. Centre National de la Recherche Scientifique, Laboratoire de Chimie des Polymères Organiques, UMR 5629, F-33607 Pessac Cedex, France

c. ITERG, 11 rue Gaspard Monge, Parc Industriel Bersol 2, F-33600 Pessac Cedex, France

Corresponding authors: taton@enscbp.fr, cramail@enscbp.fr

**Abstract**

Widely available vegetable oils were readily derivatized into chemically pure $AB_n$-type monomers (n = 2 or 3). Their polymerization led to unprecedented hyperbranched polyesters. Four different $AB_2/AB_3$-type monomers bearing one A-type methyl ester and two or three B-type alcohol functions were purposely synthesized via two elementary steps, i.e. epoxidation of the internal double bond of the vegetable oil precursors followed by ring-opening of the epoxy groups in acidic conditions. The polycondensation of these bio-sourced monomers was performed in bulk, in presence of an appropriate catalyst, giving access to modular hyperbranched polyesters of tunable properties. Among the catalysts tested, zinc acetate, 1,5,7-triazabicyclo[4.4.0]dec-5-ene (TBD) and sodium methoxide were proved the most effective, allowing to achieve molar masses in the range 3 000-10 000 g.mol$^{-1}$ and rather dispersities varying from 2 to 15, depending on the initial conditions. The degree of branching, DB, as determined by $^1$H NMR spectroscopy, was found in between 0.07 to 0.45. The as-devised hyperbranched polyesters displayed either amorphous or semi-crystalline properties, as a function of the selected $AB_2/AB_3$-type initial monomers, with a glass transition temperature, $T_g$, ranging from -33 to 9°C and a decomposition temperature at 5% wt. of the sample, $T_d^{5\%}$, varying from 204 to 340°C.

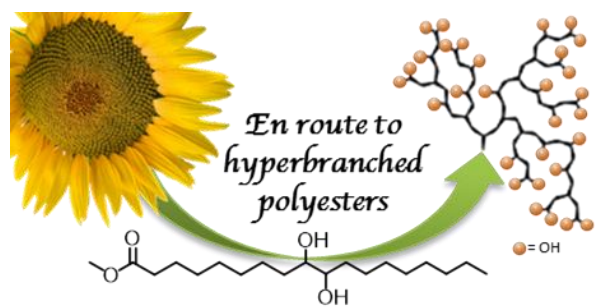

**Introduction**

The last three decades have witnessed the rapid development of hyperbranched polymers (HBPs).[1] The first intentional synthesis of HBPs dates back to 1987 with the polyphenylenes of Kim and Webster,[2] long after Flory had predicted in the 1950's that $AB_n$-type monomers should afford soluble and highly branched materials of globular shape, *via* a self-condensation polymerization process.[3] The field is still in expansion reaching more than 500 publications in recent years.[4] HBPs are a special class of dendritic materials and possess as common features with dendrimers a high branching density as well as a compact and globular architecture. This unusual structure endows them with unique properties as compared to their linear counterparts. Of particular interest are their lower viscosity both in solution and in the molten state, an improved solubility and a high functionality. Indeed, their large number of terminal functional groups offers the possibility for further modifications, providing an additional method for the variation of HBPs' properties and broadening the scope of their potential applications. In contrast to regular dendrimers,[5] HBPs are prepared in one-pot, hence they exhibit an overall structure that is not flawless. In particular, HBPs are characterized by the presence of linear units coexisting with dendritic and terminal units, which is mirrored by their degree of branching (DB) that is most often lower than unity. In addition, related one-pot synthetic methods to HBPs are usually more straightforward and more economically viable, compared to dendrimer synthesis.

Tremendous synthetic efforts have been made in the past decades to access a wide range of HBPs, from the polycondensation of $AB_n$-type monomers to less conventional polymerization methods.[4,6–11] In this context, hyperbranched polyesters have been the most investigated. This is due to the wide availability of various suitable monomers and the relative ease of synthesis of this subclass of HBPs. The success of hyperbranched polyesters (HBPEs) has resulted in the marketing of miscellaneous Boltorn[TM] products based on 2,2-bis(methylol)propionic acid (bis-MPA).[12,13] As a matter of fact, HBPEs have been nearly exclusively designed starting from fossil resources.[14–18] In a context of price volatility and uncertain supply of oil and gas, combined with environmental concerns (global warming, waste production, etc.), there is yet an urgent need to release the chemical industry from its dependence on petroleum resources.[19] In recent years, particular attention has been paid to biomass as a sustainable source of carbon.[20,21] Among available renewable resources, vegetable oils and their derivatives, *i.e.* fatty acid methyl esters (FAMEs), represent a promising feedstock for the polymer industry owing to their abundant availability, relative low cost and inherent degradability.[22–28] Interests in FAMEs are also motivated by the presence of both ester and double bonds enabling countless derivatizations and design of a wide variety of functional building blocks and related materials.

Synthesis of vegetable oil-based linear polyesters has been extensively studied in the past decade,[29–31] including by our group.[32] In contrast, only a handful of studies have been dedicated to HBPEs derived from plant oils. Related examples have involved the polycondensation of $AB_n$-type monomer precursors ($n \geq 2$). For that purpose, selective modifications/derivatizations of hydroxy-containing fatty acid derivatives have been implemented. For instance, Meier *et al.* and Li *et al.* have resorted to the thiol-ene addition of 1-thioglycerol onto methyl 10-undecenoate, as a means to access an $AB_2$-type monomer, the polycondensation of which, in presence of organic or metallic catalysts, has led to bio-based HBPEs.[33,34] Meier *et al.*[33] have also reported on the polycondensation of the same $AB_2$-type monomer using glycerol as a core molecule, and 1,5,7-triazabicyclo[4.4.0]dec-5-ene (TBD) as a catalyst at 120 °C, resulting in HBPEs with moderate molar masses ($\bar{M}_n$ = 3 500-4 400 g.mol$^{-1}$, $Đ$ = 1.87-2.85). Li *et al.*[34] have achieved higher molar masses ($\bar{M}_n$ = 11 400-60 400 g.mol$^{-1}$) though with broader dispersities ($Đ$ = 5.2-25.3), upon using a metallic catalysis based on Ti(OBu)$_4$, Sb$_2$O$_3$ or Zn(OAc)$_2$ operating at 160-170 °C. DB values determined by quantitative $^{13}$C NMR have been found in the range 0.38-0.54, in agreement with the value of 0.5 predicted by theory.[35] As for Petrović *et al.*, they have developed two synthetic pathways to HBPEs from hydroxylated FAMEs obtained either by (i) hydrogenation[36] of epoxidized soybean oil or by (ii) hydroformylation.[37,38] High molar masses (up to 42 000 g.mol$^{-1}$) have been obtained only in the latter case, with a weight average hydroxyl functionality of the polymer up to 23.2. Note, however, that no DB value has been reported. Two other reports have dealt with HBPE synthesis *via* the $A_2 + B_3$ approach, through the copolymerization of various α,ω-diacids with glycerol.[39,40]

In this contribution, we describe a platform of new $AB_n$-type building blocks based on FAMEs and the synthesis of related vegetable oil-based HBPEs with tunable properties. For that purpose, sunflower, castor and rapeseed oils were used as raw materials for the synthesis of chemically pure monomers of $AB_n$-type ($n = 2$ or 3), featuring an ester (A) and two or three alcohol (B) moieties. The polytransesterification of these monomers was optimized, by varying the initial experimental conditions. Insights into their fine structure are provided both by $^1$H NMR spectroscopy and MALDI ToF mass spectrometry. Their crystallization ability and thermal stability are also assessed by DSC and TGA analyses.

**Experimental**

**Materials.**

1,5,7-Triazabicyclo[4.4.0]dec-5-ene (TBD, 98%), meta-chloroperbenzoic acid (m-cpba, < 77%), anhydrous zinc acetate (Zn(OAc)$_2$, 99.99% trace metals basis), sodium methoxide (powder, 95%), glycerol (99%), anhydrous *tert*-butanol (ACS reagent, ≥ 99.0%), phosphoric acid (85%) and hydrogen peroxide (30%) were obtained from Sigma Aldrich. Methyl-10-undecenoate (> 96.0%) was supplied by TCI Europe. All products and solvents were used as received except otherwise mentioned. The solvents were of reagent grade quality and were purified when necessary according to methods reported in the literature. Methyl 9,10-dihydroxystearate (M2HS), methyl 9,10,12-trihydroxystearate (M3HS) and methyl 13,14-dihydroxybehenate (M2HB) were synthetized by ITERG (Pessac, France), and their synthesis is described further.

**Methods.**

*Nuclear magnetic resonance (NMR) spectroscopy.* $^1$H and $^{13}$C NMR experiments were conducted on Bruker Avance 400 spectrometer (400.20 MHz or 400.33 MHz and 100.63 MHz for $^1$H and $^{13}$C, respectively) at room temperature, in CDCl$_3$ as solvent except otherwise mentioned. Chemical shifts (δ) are reported in parts per million relative to the known solvent residual peak (δ = 7.26 ppm). DEPT-135 (Distortion Enhanced Polarization Transfer) and two-dimensional analyses such as $^1$H-$^1$H COSY (Homonuclear correlation spectroscopy), $^1$H-$^{13}$C-HSQC (Heteronuclear single quantum coherence) and $^1$H-$^{13}$C-HMBC (Heteronuclear multiple bond correlation) were also performed.

*Size exclusion chromatography (SEC) analyses* were performed in THF as the eluent (1 mL/min) at 40°C, on a PL-GPC 50 plus Integrated GPC from Polymer laboratories-Varian with a series of four columns from TOSOH [TSKgel TOSOH: HXL-L (guard column 6.0 mm ID x 4.0 cm L); G4000HXL (7.8 mm ID x 30.0 cm L); G3000HXL (7.8 mm ID x 30.0 cm L) and G2000HXL (7.8 mm ID x 30.0 cm L)]. The elution times of the filtered samples were monitored using RI detectors with a calibration curve based on low dispersity polystyrene standards (PS). Trichlorobenzene was added as flow marker.

*Differential scanning calorimetry (DSC)* measurements were carried out on DSC Q100 apparatus from TA Instruments. For each sample, two cycles from -100 to 150 °C (except otherwise mentioned) were performed at 10 °C.min$^{-1}$. Glass transition and melting temperatures were calculated based on the second heating run.

***Thermogravimetric analyses (TGA)*** were performed on two different apparatus from TA Instruments depending on the availability: TGA Q50 and Q500 at heating rate of 10 °C.min$^{-1}$ under nitrogen atmosphere from room temperature to 700 °C.

***MALDI ToF mass spectrometry (MS)*** analyses were performed by the *Centre d'Etude Structurale et d'Analyse des Molécules Organiques* (CESAMO, Bordeaux, France) on a Voyager mass spectrometer (Applied Biosystems). The instrument is equipped with a pulsed N2 laser (337 nm) and a time-delayed extracted ion source. Spectra were recorded in the positive-ion mode using the reflectron and with an accelerating voltage of 20 kV. Samples were dissolved in THF at 10 mg/ml. The IAA matrix (trans-3-indoleacrylic acid) was prepared by dissolving 10 mg in 1 ml of THF. A methanol solution of cationization agent (NaI, 10 mg/ml) was also prepared. The solutions were combined in a 10:1:1 volume ratio of matrix to sample to cationization agent. One to two microliters of the obtained solution was deposited onto the sample target and vacuum-dried.

**Synthetic procedures.**

*Synthesis of methyl 9,10-dihydroxystearate (M2HS).*

*Epoxidation step*. Methyl esters of high oleic sunflower oil (40 kg, 133.3 mol) and formic acid (1.7 kg, 37.8 mol) were added in a reactor equipped with a mechanical stirrer, a dropping funnel and a condenser. The resulting mixture was heated at 40 °C for 1 hour under stirring. Hydrogen peroxide (35%, 24.4 kg, 251.6 mol) was added dropwise to the reactor at 40 °C, using a dropping funnel to maintain the temperature in the reactor close to 70–75°C. As the reaction is exothermic, a cooling system was used to cool down the reactor. The reaction was monitored by gas chromatography and epoxide titration. The reaction mixture was then cooled down to room temperature and the aqueous phase was discarded. The organic layer was washed with an aqueous solution of sodium hydroxide (0.1 N) until the pH became neutral. The organic phase was then dried under vacuum at 60 °C to afford a clear and slightly yellow liquid.

*Hydroxylation step*. Epoxidized fatty acid methyl esters (10 kg) were placed along with an aqueous solution of phosphoric acid (12 % w/w, 5 kg) and *tert*-butanol (3 kg) as solvent in a reactor equipped with a condenser and a mechanical stirrer. The resulting mixture was heated at 90 °C under vigorous stirring. The reaction was monitored by gas chromatography. When the reaction was completed, the aqueous phase was discarded at 50°C. *tert*-Butanol was eliminated under vacuum distillation. The organic phase was then washed with hot water until the pH reached 6-7 and dried under vacuum to afford a white solid. M2HS was then recrystallized in cyclohexane (3 times, 40-60 g.L$^{-1}$) and dried under vacuum to afford a white solid powder. The product was then dissolved in a minimum of dichloromethane (DCM) and injected in a Flash chromatography apparatus from Grace. The constituents were separated on a silica column, using a dichloromethane-methanol gradient and an Evaporating Light Scattering Detector (ELSD). Two fractions were collected, corresponding to M2HS and its acid form, respectively. M2HS purity was determined by GC (97.8%).

The procedure followed to prepare methyl 13,14-dihydroxybehenate (M2HB) and methyl 9,10,12-trihydroxystearate (M3HS) was nearly identical to that described for M2HS; details are provided hereafter.

*Synthesis of methyl 13,14-dihydroxybehenate (M2HB):* methyl ester of refined erucic acid rapeseed oil. After dihydroxylation, the crude mixture was purified by means of neutralization with potassium hydroxide and recrystallization in cyclohexane, to afford M2HB as a white powder (purity: 94%). Yield of the overall synthesis: 80%.

*Methyl 9,10,11-trihydroxystearate (M3HS):* methyl ester of castor oil. M3HS was purified by recrystallization in cyclohexane followed by a flash chromatography performed on a silica column using a dichloromethane:methanol gradient (95:5). The AB$_3$-type monomer was obtained as a white powder in rather low yield (25%). Purity: 98.1%.

*Synthesis of methyl 10,11-dihydroxyundecanoate (M2HU).*

*Epoxidation step.* Methyl undecenoate (15 g, 0.076 mol) and *m*-cpba (39.2 g, 0.227 mol) were stirred at room temperature in DCM (20 mL.g$^{-1}$ of product) overnight. 3-chlorobenzoic acid formed as side product precipitated in DCM. The reaction mixture was first filtered to remove 3-chlorobenzoic acid, washed with aqueous sodium sulfite Na$_2$SO$_3$ (3 x 50 mL), aqueous sodium bicarbonate NaHCO$_3$ (4 x 50 mL) and brine (2 x 50 mL) until the pH became neutral. The organic phase was then dried over anhydrous magnesium sulfate, filtered and DCM was removed on a rotary evaporator to afford methyl 10-epoxyundecenoate. Yield: 92%.

*Hydroxylation step.* Methyl 10-epoxyundecenoate (2.5 g) was charged in a round-bottom flask equipped with a mechanical stirrer, an oil bath and a condenser. The epoxidized intermediate was dissolved in 50 mL of a 1:1 (v/v) mixture of water and *tert*-butanol under stirring. After the addition of phosphoric acid (85% w/w, 3 wt%), the reaction flask was heated under reflux at 90°C. 4 hours later, the aqueous phase was discarded at 50°C and *tert*-butanol was removed under vacuum distillation. After the addition of 50 mL DCM, the organic phase was washed twice with water (2 x 50 mL) and brine (1 x 50 mL), dried over anhydrous magnesium sulfate, filtered and the solvent was removed on a rotary evaporator. M2HU was obtained as a white powder (purity: 99%). Yield: 57%.

*Synthesis of hyperbranched polyesters.*

Polytransesterification reactions were performed in bulk, in a Schlenk flask equipped with a magnetic stirrer, a nitrogen inlet tube and an oil-bath heating system. AB$_n$-type monomers (*n* = 2 or 3) were dried at 90 °C, *i.e.* above their melting point, under dynamic vacuum prior to use. This pre-drying step took one hour after the reaction mixture was placed under nitrogen blowing. Temperature was then raised to 120 °C and the catalyst was introduced at a loading of 1.5 wt% relatively to the monomer. The mixture was allowed to react under stirring at 120 °C for 2 hours. In a second stage, the temperature was raised to $T_2$ (°C) and dynamic vacuum was applied in order to remove the released methanol. Reaction conditions were optimized for each monomer as follows. M2HS, M3HS and M2HB were preferably polymerized at $T_2$ = 160 °C during 15-24 hours and M2HU at $T_2$ = 140 °C during 8-10 hours. HBPEs prepared from M2HS and M3HS were obtained as colorless and highly viscous materials, while polycondensation of M2HB and M2HU afforded whitish waxes. The observed change in texture was the first sign of the semi-crystalline character of these HBPEs.

**Results and discussion**

**1- Synthesis of bio-based AB$_n$-type monomers**

Unsaturated FAMEs serving as raw plant oil materials were first chemically modified, so as to achieve different precursors of AB$_2$/AB$_3$-type consisting of methyl ester (A) and alcohol (B) reactive functions. The conversion of A-type FAMEs into AB$_n$-type monomers ($n$ = 2 or 3) was accomplished following a two-step straightforward procedure, involving (i) epoxidation of the double bond and (ii) subsequent ring-opening of the as-formed epoxide under acidic conditions. Derivatization of C=C double bonds into 1,2-diols through epoxidation has been widely applied to vegetable oil derivatives.[41] Upon using methyl oleate, it leads to the formation of methyl 9,10-dihydroxystearate (M2HS), *i.e.* an AB$_2$-type monomer (Scheme 1). Other substrates of interest were also considered, on the basis of the same strategy. Thus, methyl euricate, a C22-carbon atom-containing precursor (instead of C18 for methyl oleate), allowed accessing methyl 13,14-dihydroxybehenate (M2HB). The latter monomer was expected to provide HBPEs with a higher alkyl chain length between branching points compared to M2HS (11 CH$_2$ vs. 7 CH$_2$). Similar modification of methyl ricinoleate, where a hydroxyl function is already present, yielded the AB$_3$-type monomer precursor denoted as M3HS. Lastly, methyl undecenoate (C11:1) was selected to prepare methyl 10,11-dihydroxyundecanoate (M2HU), an AB$_2$-type monomer, the polycondensation of which should lead to HBPEs free of dangling chains.

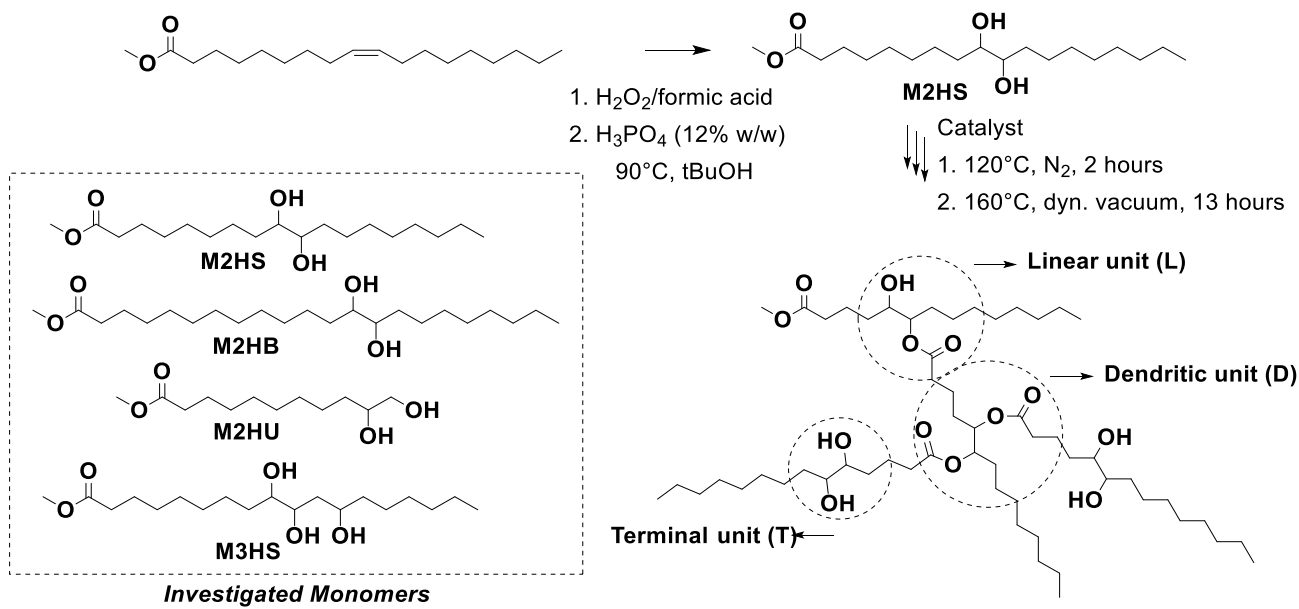

**Scheme 1.** Synthetic route to novel AB$_n$-type fatty acid-based monomers and related hyperbranched polyesters.

All monomer precursors were obtained as white powders in yields higher than 60%, except for M3HS (≈ 25%). The low yield in the latter case could be explained by the hydroxylation of methyl ricinoleate which resulted in a mixture of four diastereoisomers, only one of them being able to crystallize.[42–45] The structure of the so-formed AB$_2$- or AB$_3$-type monomers was assessed by $^1$H NMR spectroscopy (see ESI Figure S1). GC analyses revealed a chemical purity higher than 94% in all cases.

## 2- Synthesis of hyperbranched polyesters by self-polycondensation of AB$_2$- and AB$_3$–type monomers

All polymerizations were carried out in bulk, in line with many of the *Principles of Green Chemistry*,[46] following a two-step-one-pot methodology illustrated in Scheme 1. After a two-hour oligomerization stage at 120 °C, the temperature was raised to 160 °C, and dynamic vacuum was applied. Polycondensations were thus performed in the melt, and crude polymers were analyzed without further purification (**Table I**).

**Table I.** Screening of various catalysts for the step-growth polymerization of M2HS

| Entry | Catalyst | $x^a$ (%) | $\bar{M}_n^a$ (g.mol$^{-1}$) | $Đ^a$ |
|---|---|---|---|---|
| 1 | Ti(OBu)$_4$ | 44 | 1 100 | 1.09 |
| 2 | Ti(OiPr)$_4$ | 48 | 1 200 | 1.08 |
| 3 | Zn(OAc)$_2$ | 98 | 3 500 | 2.71 |
| 4 | DBTO | 39 | 1 100 | 1.09 |
| 5 | Sb$_2$O$_3$ | 45 | 1 200 | 1.08 |
| 6 | TBD | 100 | 4 100 | 2.46 |
| 7 | *m*-TBD | 21 | 1 000 | 1.10 |
| 8 | DABCO | No polymerization | | |
| 9 | DBU | No polymerization | | |
| 10 | NaOMe | 100 | 6 100 | 3.08 |

(a) determined by SEC in THF; calibration with PS standards. Procedure: 2 hours at 120 °C under N$_2$ followed by 13 hours at 160 °C under dynamic vacuum

Various commercially available transesterification catalysts were screened with a loading of 1.5 wt.% relatively to the monomer. A more systematic investigation was undertaken with M2HS (**Table I**). Because transition metal catalysts currently dominate condensation polymerization processes at the industrial scale, a selection of zinc, tin, titanium and antimony-based catalysts was considered first. Organic catalysts were also investigated in line with the development of environmentally friendly metal-free processes.[47,48] Guanidines including 1,5,7-triazabicyclo[4.4.0]dec-5-ene (TBD), its *N*-methyl derivative, namely, 7-methyl-1,5,7-triazabicyclo[4.4.0]dec-5-ene (*m*-TBD), the amidine base 1,8-diazabicyclo[5.4.0]undec-7-ene (DBU) and the tertiary amine 1,4-diazabicyclo[2.2.2]octane (DABCO) were thus examined. Finally, sodium methoxide (NaOMe) was also tested as it is widely used in oleochemistry as transesterification catalyst of crude vegetable oils.[49]

Most of the organometallic catalysts tested, including titanium (IV) butoxide and isopropoxide, $Ti(OBu)_4$ and $Ti(OiPr)_4$, respectively, dibutyltin(IV) oxide (DBTO) and antimony trioxide ($Sb_2O_3$), along with the organic base *m*-TBD showed a poor efficiency toward the polycondensation of M2HS. DABCO and DBU did not show any activity at all. Much better results were achieved with zinc acetate, $Zn(OAc)_2$, TBD and NaOMe. Based on this catalyst screening, $Zn(OAc)_2$, TBD and NaOMe were selected for further polymerization/optimization utilizing the other $AB_n$-type bio-sourced monomers.

**Table II.** Molar masses, dispersity and thermal properties of HBPEs obtained by self-polytransesterification of M2HS, M2HB, M2HU and M3HS

| Entry | Monomer | Catalyst | $T^a$ (°C) | $M_n^b$ (g.mol$^{-1}$) | $Đ^b$ | $DB_{Frey}^c$ | $T_g^d$ (°C) | $T_m^d$ (°C) | $T_c^d$ (°C) | $T_d^e$ 5wt.% (°C) |
|---|---|---|---|---|---|---|---|---|---|---|
| P1 | M2HS | Zn(OAc)$_2$ | 160 | 3 500 | 2.71 | 0.07 | -27 | - | - | 246 |
| P2 | | Zn(OAc)$_2$ | 170 | 6 300 | >15 | 0.26 | -30 | - | - | 252 |
| P3 | | TBD | 160 | 4 100 | 2.46 | 0.35 | -24 | - | - | 320 |
| P4 | | TBD | 170 | 5 300 | 3 | 0.38 | -24 | - | - | 324 |
| P5 | | TBD$^6$ | 170 | 7 600 | >12 | 0.41 | -22 | - | - | 338 |
| P6 | | NaOMe | 160 | 6 100 | 3.08 | 0.29 | -20 | - | - | 340 |
| P7 | M2HB | Zn(OAc)$_2$ | 160 | 3 000 | 1.93 | 0.09 | - | nd. | nd. | nd. |
| P8 | | Zn(OAc)$_2^f$ | 160 | 5 600 | >11 | 0.18 | - | 20 | 14 | 260 |
| P9 | | TBD | 160 | 5 600 | 3.05 | 0.33 | - | 22 | 12 | 298 |
| P10 | | NaOMe | 160 | 9 200 | 3.27 | 0.30 | - | 23 | 12 | 332 |
| P11 | M2HU | Zn(OAc)$_2$ | 140 | 2 400 | 2.33 | nd. | - | 61 | 44 | 204 |
| P12 | | TBD | 140 | 2 400 | 2.67 | 0.25 | - | 54 | 36 | 230 |
| P13 | M3HS | Zn(OAc)$_2$ | 160 | 3 800 | 2.58 | nd. | -3 | - | - | 254 |

| | | | | | | | | | |
|---|---|---|---|---|---|---|---|---|---|
| P14 | | TBD | 160 | 5 600 | 4.05 | nd. | -1 | - | - | 305 |
| P15 | | NaOMe | 160 | 4 600 | 2.83 | nd. | 3 | - | - | 299 |

(a) Temperature of polymerization. (b) SEC in THF - calibration PS standards. (c) $^1$H NMR. (d) DSC – 10 °C.min$^{-1}$. (e) determined by TGA – 10 °C.min$^{-1}$ under N$_2$. (f) Duration of polymerization increased from 15 to 24 hours. nd. = not determined.

As summarized in Table II , M2HS, M3HS and M2HB were efficiently polymerized at 160 °C, monomer conversion over 95% being reached within 15 hours. SEC analyses indicated the formation of HBPEs with molar masses (Mn) in between 3 000 and 6 100 g.mol-1, corresponding to an average number degree of polymerization ($\overline{DP_n}$) ranging from 10 to 12. These molar masses were determined from linear PS standards, and thus should be taken with caution, since HBPEs likely display much smaller hydrodynamic radii than that of a linear homologue or of a PS standard of same molar mass. Dispersity values higher than 2 were obtained, as expected for hyperbranched materials prepared by polycondensation of ABn-type monomers.50 Higher molar masses, Mn = 7 600 g.mol-1, were reached by increasing either the reaction time, from 15 to 24 hours, and/or the reaction temperature up to 170 °C, as highlighted with M2HS and M2HB (Table II, P1 vs. P2, P3 vs. P4 and P5, P7 vs. P8). The impact was even more important on dispersities, significantly broader values being obtained (> 10) in agreement with the theory. However, higher polymerization temperatures (≥ 200 °C) and longer reaction times (over 40 hours) were found to favor the formation of insoluble gel-like materials.

In contrast to the other bio-based monomers, M2HU was preferably polymerized at 140 °C. Due to the higher reactivity of its primary alcohols, it was indeed noted that gelation occurred more rapidly at 160 °C, *i.e.* in less than 10 hours. Yet, further decrease of the polymerization temperature to 120 °C yielded oligomers only (M$_n$ < 900 g.mol$^{-1}$). Within 8 hours at 140 °C, full conversions were achieved using both zinc acetate and TBD as catalysts, leading to molar mass around 2 500 g.mol$^{-1}$.

All these experiments revealed the importance of suitable experimental conditions so as to achieve soluble HBPEs. According to Flory's theory, the statistical polycondensation of AB$_n$-type monomers yields HBPs without any risk of gelation provided that (i) A reacts exclusively with B and (ii) reactions do not involve internal cyclization.[3] Though, in practice, side reactions can take place during HBP synthesis and can also be responsible for the formation of insoluble material due to cross-linking. Syntheses involving M2HS were further scrutinized by analyzing M2HS-based HBPEs by MALDI-ToF mass spectroscopy (Figure 1).

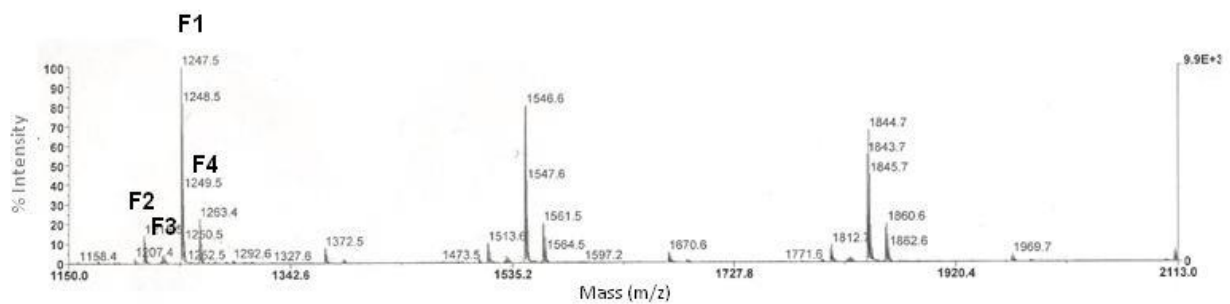

**Figure 1.** MALDI-ToF MS analysis of M2HS-based HBPE (Matrix: trans-3-idoleacrylic acid, entry P3, Table II)

Four distinct families of populations were eventually detected with an expected peak-to-peak mass increment of 298.1 g.mol$^{-1}$ matching the repeating unit. Population F1 corresponds to the targeted HBPE chains derived from M2HS, *i.e.* n x ($M_{unit}$) + $M_{MeOH}$ + $M_{Na}$ where $M_{unit}$ = $M_{M2HS}$ - $M_{MeOH}$ = 298.1 g.mol$^{-1}$. Population F2 shows a mass difference of -32 g.mol$^{-1}$, compared to the former regular HBPE structure. This was ascribed to the presence of lactone units (n x ($M_{unit}$) + $M_{Na}$), as the result of an esterification reaction occurring by an intramolecular pathway. The latter side reaction could either involve the carbonyl group of the focal point and a hydroxyl function of the same HBPE molecule (F2, Scheme 2), or could take place *via* a hydroxyl-ester interchange reaction between groups of the same branch (F'2, **Scheme 2**). As for population F3, it was characterized by $m/z$-18 g.mol$^{-1}$ with regards to F1 (**Scheme 2**), and it could be explained by a loss of a water molecule. In other words, this population reflected the occurrence of intramolecular etherification between two hydroxyl groups: n x ($M_{unit}$) + $M_{MeOH}$ + $M_{Na}$ - $M_{H2O}$. It is noteworthy that, if intramolecular cyclizations through ester (F2) or ether (F3) bond formation might reduce achievable molar masses of these bio-based HBPEs, they did not yield cross-linked materials. Gelation observed experimentally would thus be due to extensive etherification side reactions taking place intermolecularly, and thus leading to more than one focal group per macromolecule, *i.e.* at n x ($M_{unit}$) + 2x $M_{MeOH}$ + $M_{Na}$ -$M_{H2O}$. Accordingly, a series of peaks (= population F4) was observed, at $m/z$+14 g.mol$^{-1}$ (= $M_{MeOH}$ - $M_{H2O}$), and assigned to ether bridges between polymeric chains (**Scheme 3**).

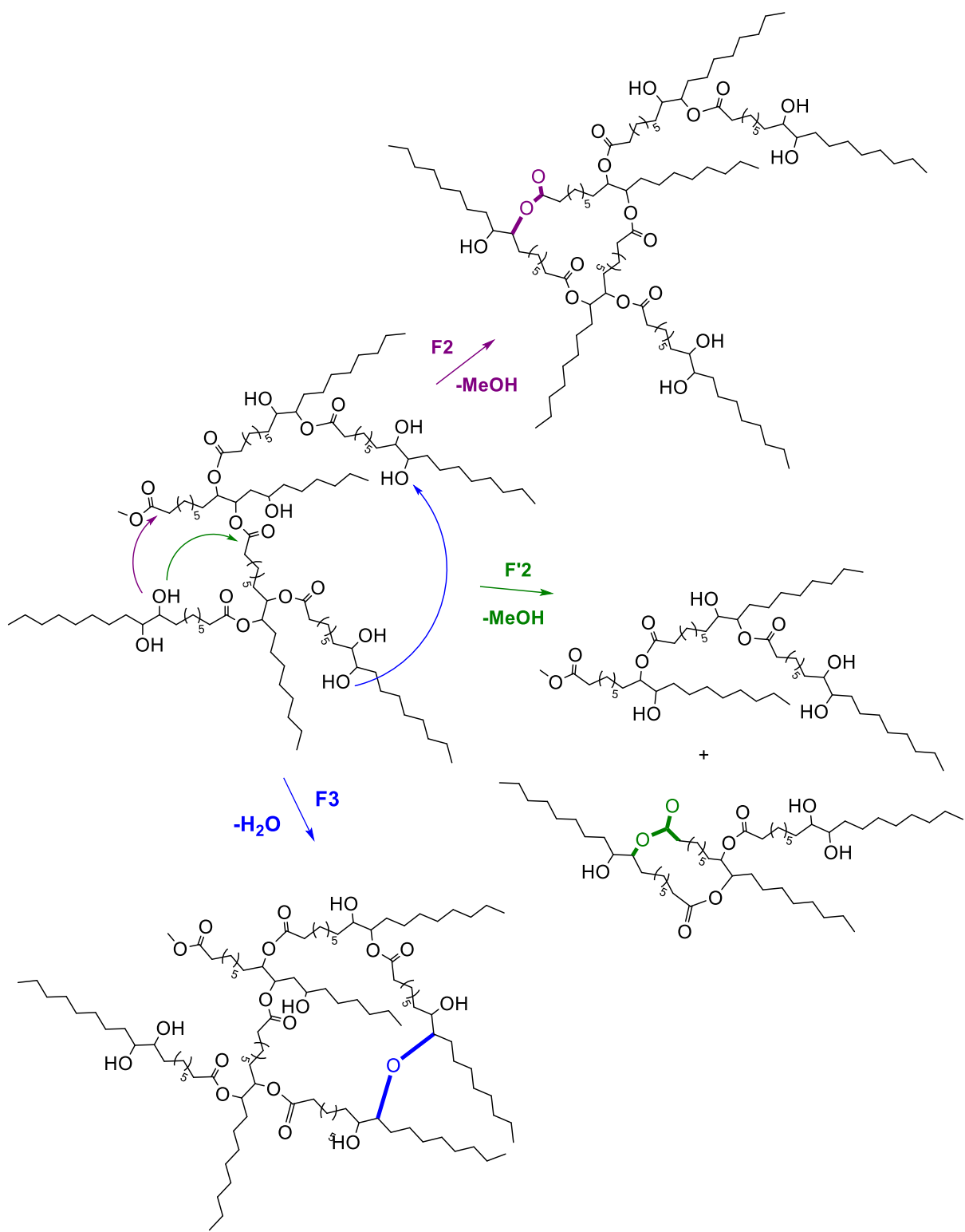

**Scheme 2.** Proposed side reactions during bio-based HBPE synthesis and related populations: cyclization by intramolecular esterification (F2), by intramolecular hydroxyl-ester interchange between groups of the same branch (F'2) and by intramolecular etherification (F3).

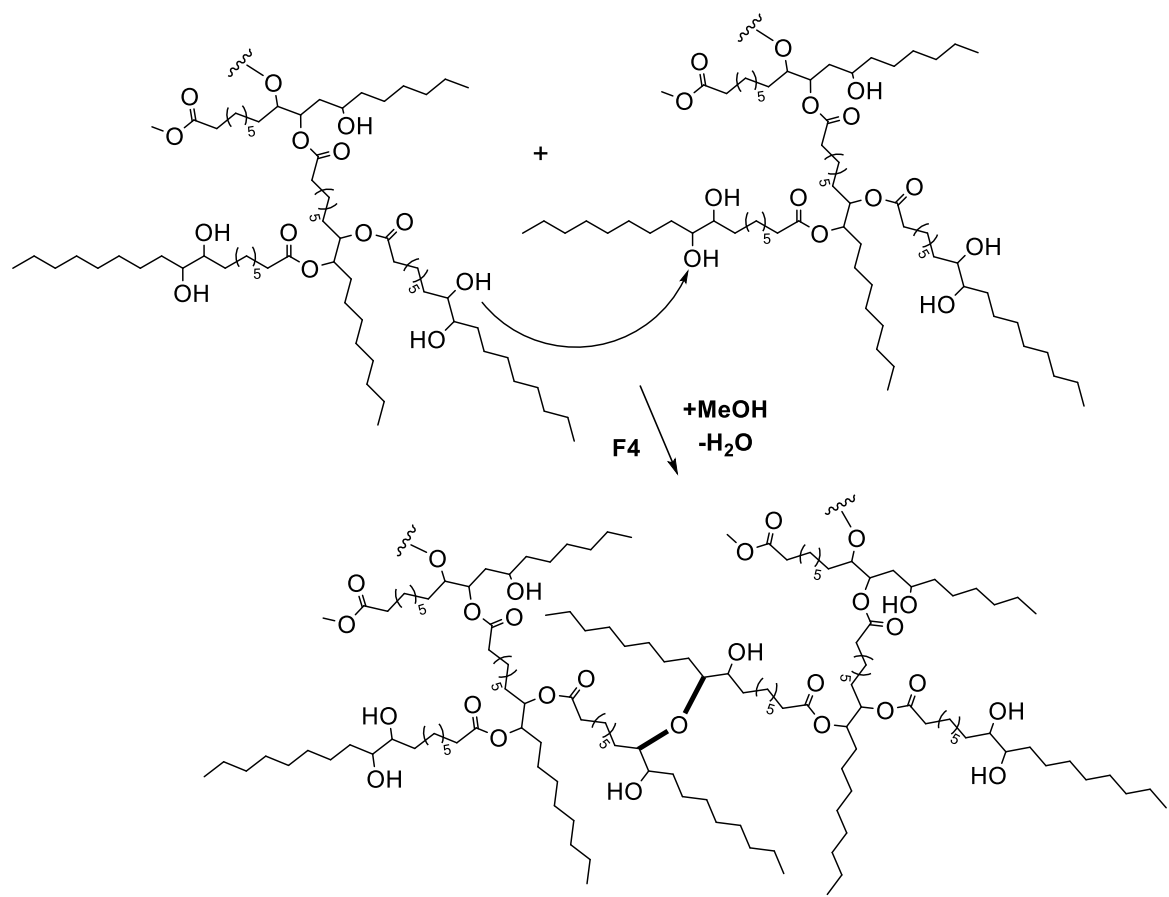

**Scheme 3.** Intermolecular etherification accounting both for the presence of population F4 and for gel formation.

Although these side reactions could not be quantified, they appeared to occur whatever the polymerization catalyst employed (Zn(OAc)$_2$ and TBD), and thus compete with the transesterification reaction during the whole polymerization process. These observations were supported by SEC analyses, as illustrated in **Figure 2**, comparing SEC-RI traces of aliquots of the TBD-catalyzed polycondensation of M2HS. In the early stages of reaction, indeed, chromatograms displayed the pattern of a typical step-growth polymerization of AB$_n$-type monomers. Observation of multimodal distributions was consistent with the formation of intermediate oligomers of increasing degrees of polymerization ($\overline{DP_n}$). The disappearance of the peak at 27.9 minutes, assigned to M2HS, evidenced a complete conversion within the first 4 hours of polymerization. After 15 hours of reaction, signals attributed to dimers and trimers at 26.4 and 25.3 minutes, respectively, were split into two discrete peaks, as the result of lactone formation by cyclization, as discussed above. In absence of focal point constituted of a methyl ester function, these cyclic oligomers cannot react further during polymerization and remain as such till the last stage of the reaction, thereby contributing to the broadening of the molar mass distribution. Beyond 20 hours of polymerization, a shoulder was observed in the higher molar mass region, likely as the consequence of both intermolecular transesterification and etherification side reactions. This intermolecular etherification creates a second focal point in the polymer chain and can ultimately lead to crosslinking.

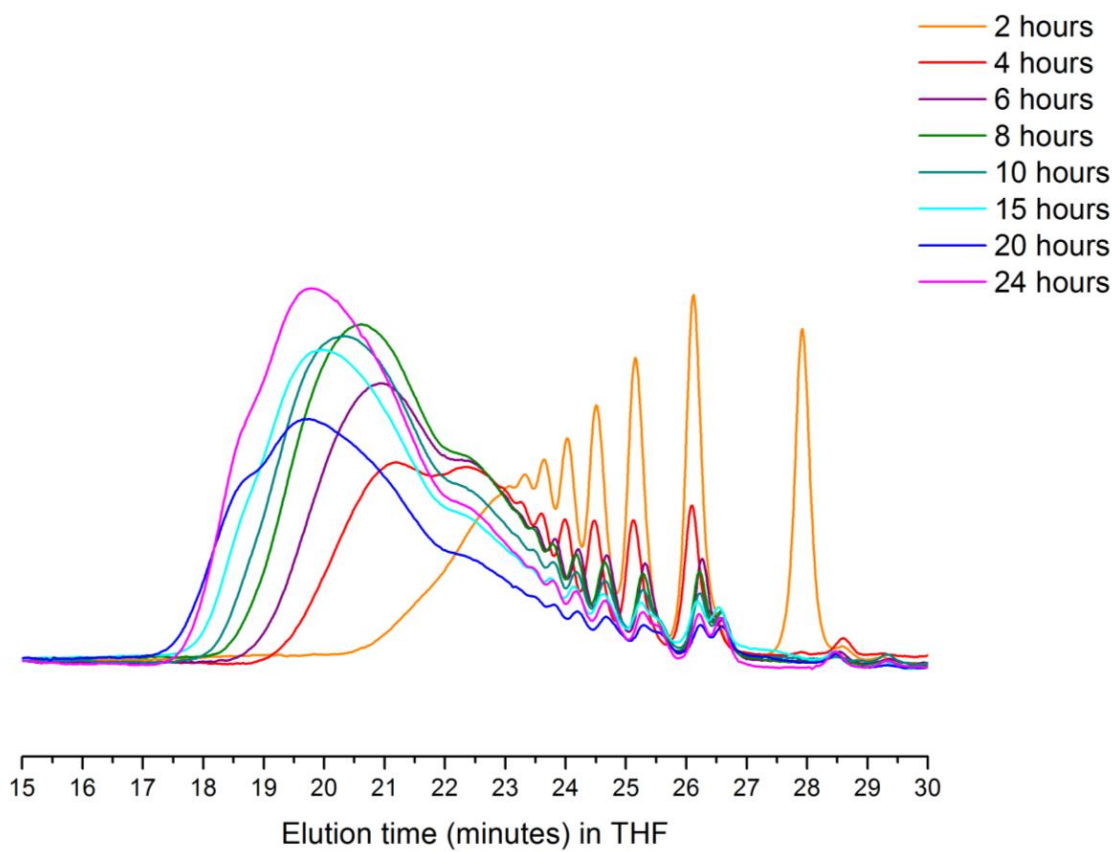

**Figure 2.** SEC-RI traces of aliquots withdrawn at different reaction times, during the TBD-catalyzed polycondensation of M2HS.

**Characterization of HBPEs**

The chemical structure of the as-formed HBPEs was first confirmed by FT-IR analysis (see ESI Figure S3). In all spectra, absorption bands characteristic of both the ester and the hydroxyl functions were observed at 1 734 and 3 400 cm$^{-1}$, respectively. To gain a better insight into their fine structure, and in particular to determine the DB value of these HBPEs, analysis by $^1$H NMR spectroscopy was realized. Fréchet et al.[51] first expressed the DB as follows:

$$DB_{Fréchet} = \frac{[D] + [T]}{[D] + [T] + [L]} \quad (1)$$

where [D], [T] and [L] correspond, respectively, to the molar proportion of dendritic, terminal and linear units. DB is thus equal to 1 for perfectly branched structures (e.g. for dendrimers), whereas HBP usually exhibit DB values lower than 1. Assuming that this equation is only valid for high molar mass HBPs, Frey et al.[35] extended equation (1) and proposed the following expression for DB:

$$DB_{Frey} = \frac{2[D]}{2[D] + [L]} \quad (2)$$

Both definitions lead to similar results at high conversions. In the present study, equation (2) was used for the determination of DB.

Importantly, well-separated signals assigned to the three different subunits, T, L and D, could be clearly distinguished by 1H NMR spectroscopy, allowing the DB value to be accurately determined. As an illustration, Figure 3 (see also Fig. S4 in ESI) depicts the 1H NMR spectrum of P3 of Table II . The steep decrease in intensity of the peak assigned to the methoxy group at 3.66 ppm confirmed the polyester synthesis. Consistently with the 1H NMR spectrum of M2HS, the peak at 3.38ppm was attributed to the terminal unit (T), i.e. to the protons adjacent to the vicinal diol on α-carbon atoms (Ha). After the first substitution, protons Hb and Hc, which are characteristic of the linear unit (L), are no longer equivalent and thus display down-fielded signals at 3.56 and 4.81 ppm, respectively. It should be noted that chemical shifts are equivalent after esterifying either the 9 or the 10-position. Finally, the peak at 4.99 ppm was unambiguously assigned to the protons Hd of the dendritic unit (D).

Similarly, HBPEs derived from M2HB and M2HU were successfully characterized by $^1$H NMR spectroscopy (see ESI, Figures S5-6 and Table S1). In the case of the AB$_3$-type monomer, however, as-formed HBPEs are not characterized by three but by four different units, including the terminal (T), the linear (L), the semi-dendritic (sD) and the perfectly dendritic (D) units. Since the three hydroxyl groups of M3HS are not chemically equivalent, the number of accessible configurations is increased to 8, leading to a complex NMR spectrum (see ESI, Fig. S7) that did not allow us to determine the DB value in this case.

The DB was found to dramatically depend on the type of catalyst employed with both AB$_2$-type monomers investigated in this work. While zinc acetate appeared to favor the formation of weakly branched structures (DB < 0.26), both TBD and NaOMe gave HBPEs with a DB up to 0.45 (see ESI, Figure S8).

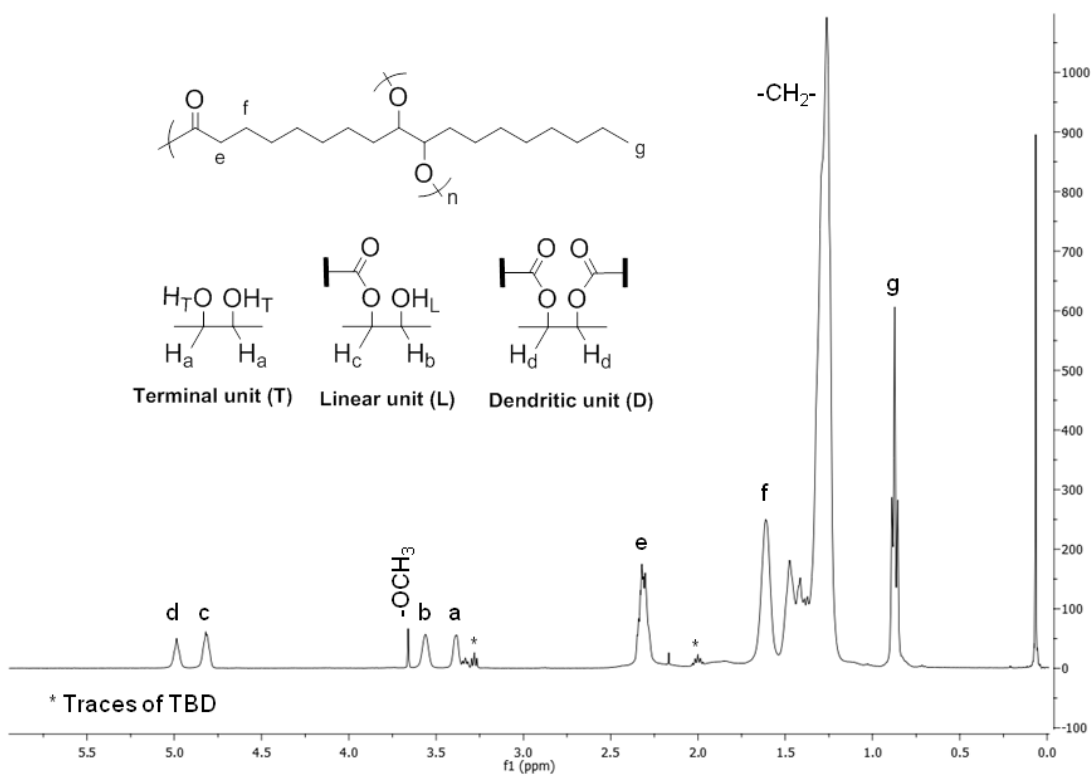

**Figure 3.** $^1$H NMR spectrum of P3 prepared by polycondensation of M2HS in CDCl$_3$ (Table II)

We also attempted to determine the DB value as a function of the conversion, on the basis of previous works by Hölter and Frey[35] and by Yan, Müller et al,[52] who expressed the conversion dependence of the DB as follows:

$$DB = \frac{2x}{5-x}\bigg|_{k_L=k_D} \quad (3)$$

where $x$ corresponds to the conversion of the focal point A. This equation describes the ideal random polycondensation of $AB_2$-type monomers during which the formation of linear ($k_L$) and dendritic ($k_D$) units takes place at equal reaction rates ($k_L = k_D$, Scheme 4). Overall, DB values were found in the range 0.07 to 0.45 (Table II), which remained lower than the value of 0.5 predicted by theory at full conversion. This somehow limited value of DB can be easily correlated to the position of the two alcohol groups of our bio-based $AB_2$-type monomers. Once a linear unit is formed, the reactivity of the remaining (non-reacted) alcohol ($-OH_L$) is indeed significantly reduced due to steric hindrance, limiting the formation of dendritic units. This phenomenon has been called by Galina et al.[53] as the negative substitution effect, on the basis of kinetic investigations.

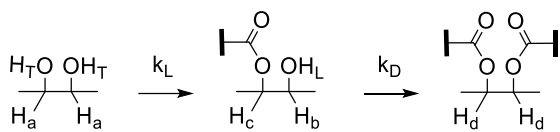

**Scheme 4.** Rate constants of the formation of linear and dendritic units: $k_L$ and $k_D$

Here, both the $Zn(OAc)_2$ and the TBD-catalyzed polytransesterifications of M2HS were monitored by $^1H$ NMR spectroscopy. Theoretical and experimental data are compared in Figure 4. The discontinuity of the curves showing the evolution of DB with the ester group conversion ($x$) is explained by the fact that the first three data were collected during the oligomerization stage at 120 °C (see above), while subsequent data were obtained when the polymerization took place at 160 °C under dynamic vacuum. This curve clearly indicates that the two hydroxyl functions of M2HS do not exhibit the same reactivity, irrespective of the catalyst, experimental points always remaining below the theoretical line. The maximum DB value of 0.4 with TBD and of 0.2 with Zn(OAc)$_2$ denotes a reactivity ratio $r$ ($r = k_L/k_D$) equal to 0.5 and to 0.1, respectively.[53]

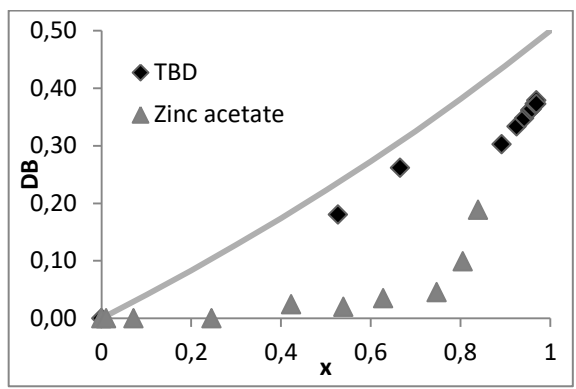

**Figure 4.** Conversion dependence of the degree of branching (DB) for the polycondensation of M2HS using TBD or zinc acetate as catalyst (solid lines: theoretical data, symbols: NMR data)

**Thermal properties**

Thermal properties of our bio-based HBPEs were assessed by DSC analysis. Glass transition and melting temperatures were thus determined from the second heating scan at 10 °C.min$^{-1}$ and crystallization temperatures from the first cooling scan at the same rate (Table II). As expected, all HBPEs derived from M2HS and M3HS proved amorphous. Compounds obtained from M3HS showed higher glass transition temperatures, *i.e.* in between -10 and 9 °C, compared to M2HS-based polyesters (-32.5 ≤ T$_g$ ≤ -20 °C), which was obviously explained by the higher functionality of the AB$_3$-type monomer (M3HS *vs.* M2HS). Unlike linear polymers (see ESI Fig. S9), the T$_g$ value of HBPs is greatly affected both by the nature and the content of end-groups (T units).[11] The number of polar end-groups increasing with the monomer functionality, more interactions likely develop within M3HS-based HBPEs, compared to M2HS-based homologues of similar molar masses. Note however that the branching density of these renewable HBPEs did not seem to impact $T_g$ values.

Interestingly, both HBPEs derived from M2HB and M2HU displayed semi-crystalline properties. DSC thermograms are given in ESI (Fig. S10). The higher flexibility of M2HB-derived HBPEs, which consist of a higher chain-length between branching points compared to M2HS (11 carbon atoms *vs.* 7), allowed the development of crystalline zones. The semi-crystalline character of M2HB-based HBPEs was confirmed by wide-angle X-ray diffraction (WAXS) measurements (see ESI; Fig. S11). The absence of pendant alkyl chains acting as plasticizers also appeared to endow the HBPEs derived from M2HU with semi-crystalline properties. In this case, complex melting and crystallization patterns were observed, probably due to the unsymmetrical nature of the initial diol that could affect the chain mobility and prevent the formation of stable crystals.

Lastly, thermal stabilities of these HBPEs were investigated by TGA under non-oxidative conditions, and at a heating rate of 10 °C.min$^{-1}$ (Table II). Typical degradation profiles are given in ESI (Fig. S12-13). HBPEs thus showed typical thermal stabilities for oily-derived polymers,[32] with a 5% weight loss ($T_{d5\%}$) up to 332 °C. HBPEs derived from M2HU were found to exhibit the lowest heat resistance of this series, likely due to their lower molar masses ($\bar{M}_n \leq 2\ 500$ g.mol$^{-1}$). Overall, HBPEs prepared using zinc acetate (P1, P2, P7, P8, P11 and P13) appeared less stable than their homologues obtained with TBD or NaOMe. The thermal degradation of the former compounds was found to be accelerated at high temperatures, likely due to residual traces of Zn(OAc)$_2$ catalyzing their depolymerization (see ESI Fig. S14), as already reported for more conventional polyesters, such as poly(ethyleneterephthalate)[54] or poly(tetramethylene succinate).[55]

**Conclusion**

Sunflower, castor and rapeseed oils can serve as raw materials to be readily derivatized into $AB_2$- or $AB_3$-type monomer substrates featuring methyl ester and alcohol functions, by means of epoxidation followed by ring-opening of epoxy-rings. Subsequent polycondensation of these bio-based monomers can be performed in bulk, giving access to modular hyperbranched polyesters through repeated transesterification reactions. Zinc acetate, TBD and sodium methoxide -three catalytic systems operating by a specific activation mechanism-, are the most effective, achieving molar masses in the range 3 000-10 000 g.mol$^{-1}$ and dispersities varying from 2 to 15. Depending on the initial conditions and on the monomer considered, either amorphous or semi-crystalline hyperbranched polyesters, with a $T_g$ value in the range -33 °C to 9 °C, a thermal stability above 300 °C, and with a degree of branching that could be varied from 0.07 to 0.45, can thus be synthesized. Due to the presence of numerous hydroxyl groups both in linear and terminal units, such hyperbranched polyesters are likely amenable to facile post-polymerization modification, allowing for further tuning the properties of these renewable branched materials.

**Funding Sources**

This work was performed, in partnership with the SAS PIVERT, within the frame of the French Institute for the Energy Transition (Institut pour la Transition Energétique (ITE) P.I.V.E.R.T. (www.institut-pivert.com) selected as an Investment for the Future ("Investissements d'Avenir"). This work was supported, as part of the Investments for the Future, by the French Government under the reference ANR-001-01.

**Acknowledgments**

The authors thank University of Bordeaux, Bordeaux INP, CNRS, Aquitaine Regional Council, the SAS PIVERT and ITERG for the support of this research. Besides, the writers acknowledge P. Castel and C. Absalon from *CESAMO* for MALDI-ToF MS experiments and A.Bentaleb from *CRPP* for the WAXS measurements.

**Electronic Supporting Information:**

**Hyperbranched Polyesters by Polycondensation**

**of Fatty Acid-based AB$_n$-type Monomers**

Blandine Testud,[a,b] Didier Pintori,[c] Etienne Grau,[a,b] Daniel Taton[a,b] [*] and Henri Cramail[a,b] [*]

a. Université de Bordeaux, Laboratoire de Chimie des Polymères Organiques, UMR 5629, Bordeaux INP/ENSCBP, 16 avenue Pey-Berland, F-33607 Pessac Cedex, France

b. Centre National de la Recherche Scientifique, Laboratoire de Chimie des Polymères Organiques, UMR 5629, F-33607 Pessac Cedex, France

c. ITERG, 11 rue Gaspard Monge, Parc Industriel Bersol 2, F-33600 Pessac Cedex, France

Corresponding authors: taton@enscbp.fr, cramail@enscbp.fr

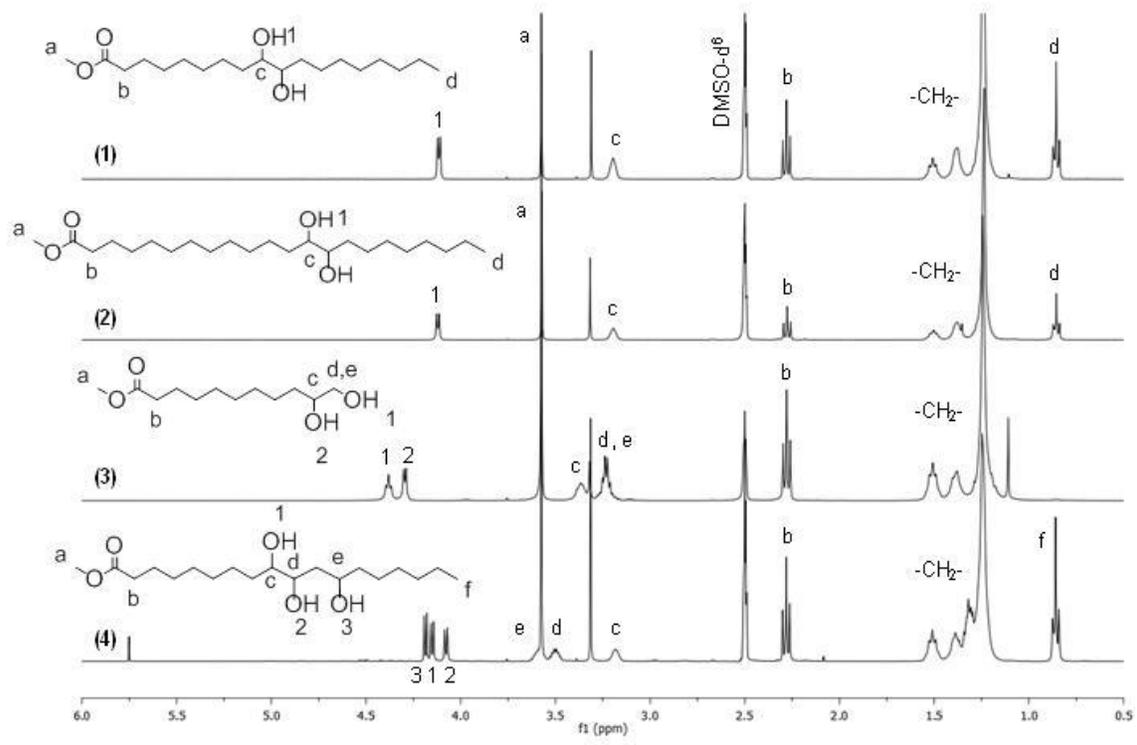

Figure SI 1. Stacked $^1$H NMR spectra of (1) M2HS, (2) M2HB, (3) M2HU and (4) M3HS in DMSO-$d_6$

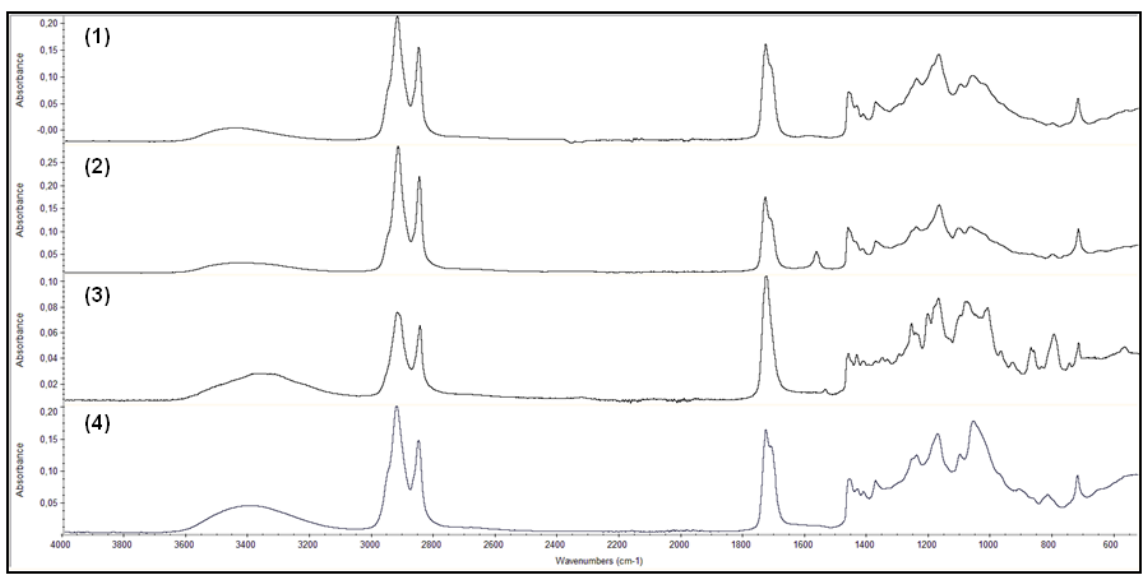

Figure SI 2. Stacked FT-IR spectra of HBPEs prepared by polycondensation of (1) M2HS, (2) M2HB, (3) M2HU and (4) M3HS

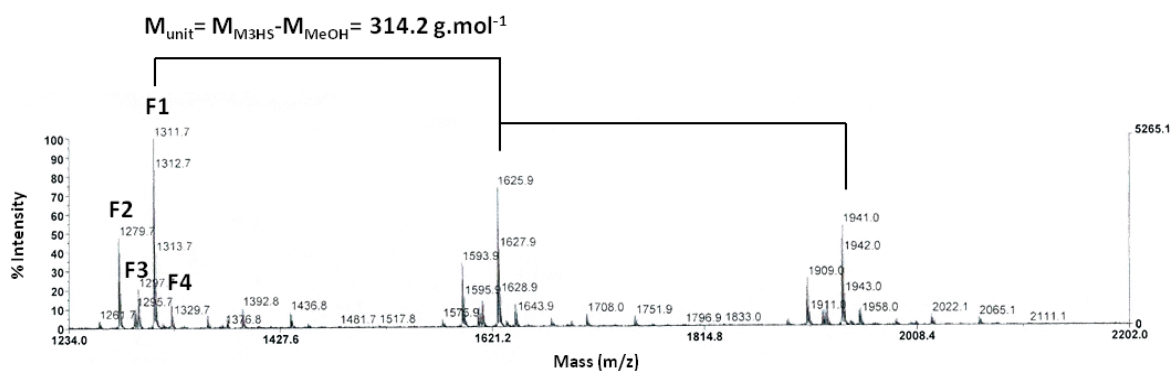

Figure SI 3. MALDI-TOF MS analysis of M3HS-based HBPE (Matrix trans-3-idoleacrylic acid)

Same nomenclature was used

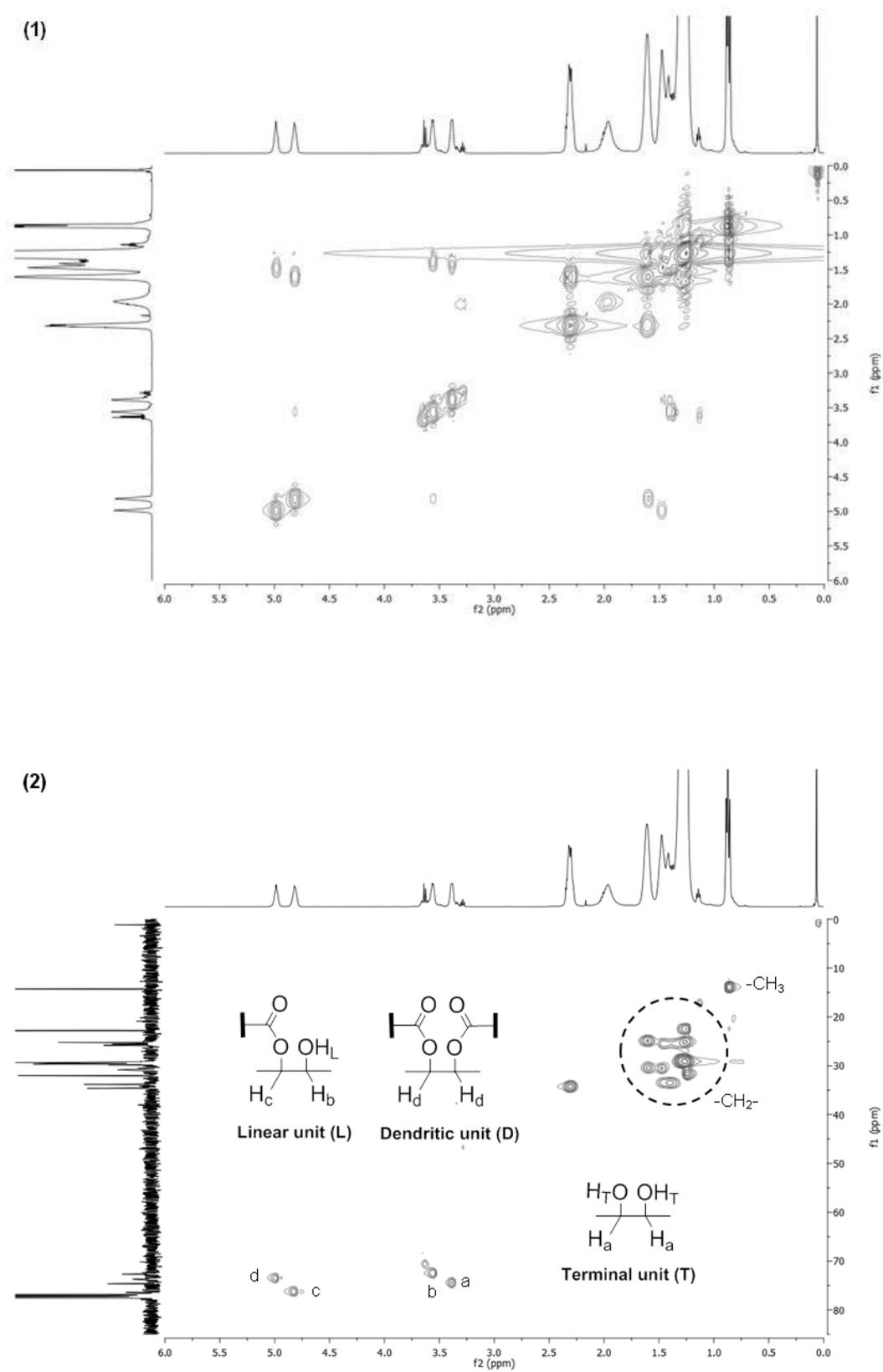

Figure SI 4. (1) $^1H$-$^1H$ COSY and (2) $^1H$-$^{13}C$ HSQC of a M2HS-based HBPE in $CDCl_3$

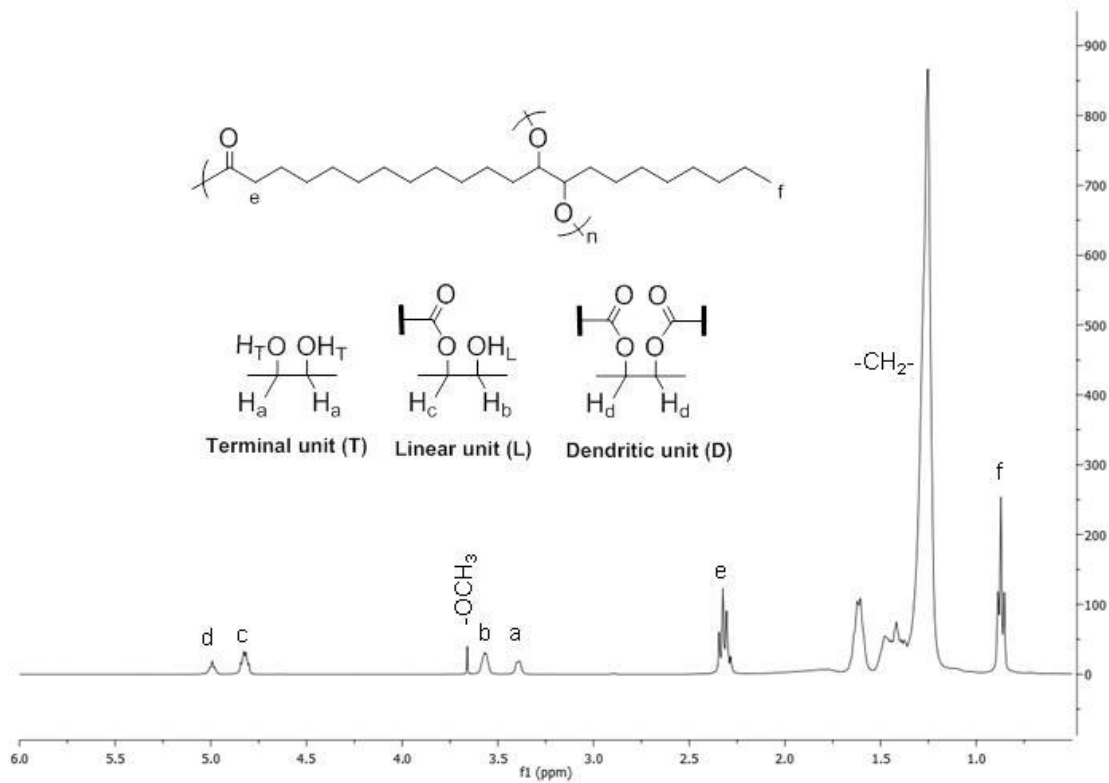

Figure SI 5. $^1$H NMR spectrum of a HBPE prepared by polycondensation of M2HB in CDCl$_3$ (P10)

Considering the unequal reactivity of the primary and secondary alcohols of M2HU, HBPEs derived from this AB2-type monomer are not described on the basis of three, but four different subunits as followed. These polyesters were preferably characterized in DMSO-d$_6$, due to a strong overlapping of the signals in CDCl$_3$. With the help of 2D NMR techniques, $^1$H NMR spectrum of P12 was partly elucidated as can be seen in Figure SI 6(1).

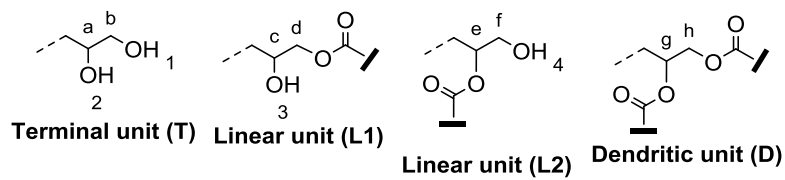

Protons $H_d$, $H_e$ and $H_g$ characteristic of the subunits L1, L2 and D, respectively, were identified at 3.85, 4.72 and 4.93 ppm. For the purpose of subsequently assessing the molar ratio in linear unit L2, a few drops of water were added in the NMR tube in order to 'switch off' the signals assigned to the hydroxyl functions. Accordingly, as can be seen in Figure SI 6 (2), all peaks attributed to $-OH_1$, $-OH_2$, $-OH_3$ and $-OH_4$ were no longer observed. Therefore, based on the respective integration of protons $H_d$ (L1), $H_e$ (L2) and $H_g$ (D), the degree of branching of P12 was found to be equal to 0.25. This result tended to indicate the formation a rather linear structure, in agreement with the higher reactivity of primary alcohols upon secondary ones.

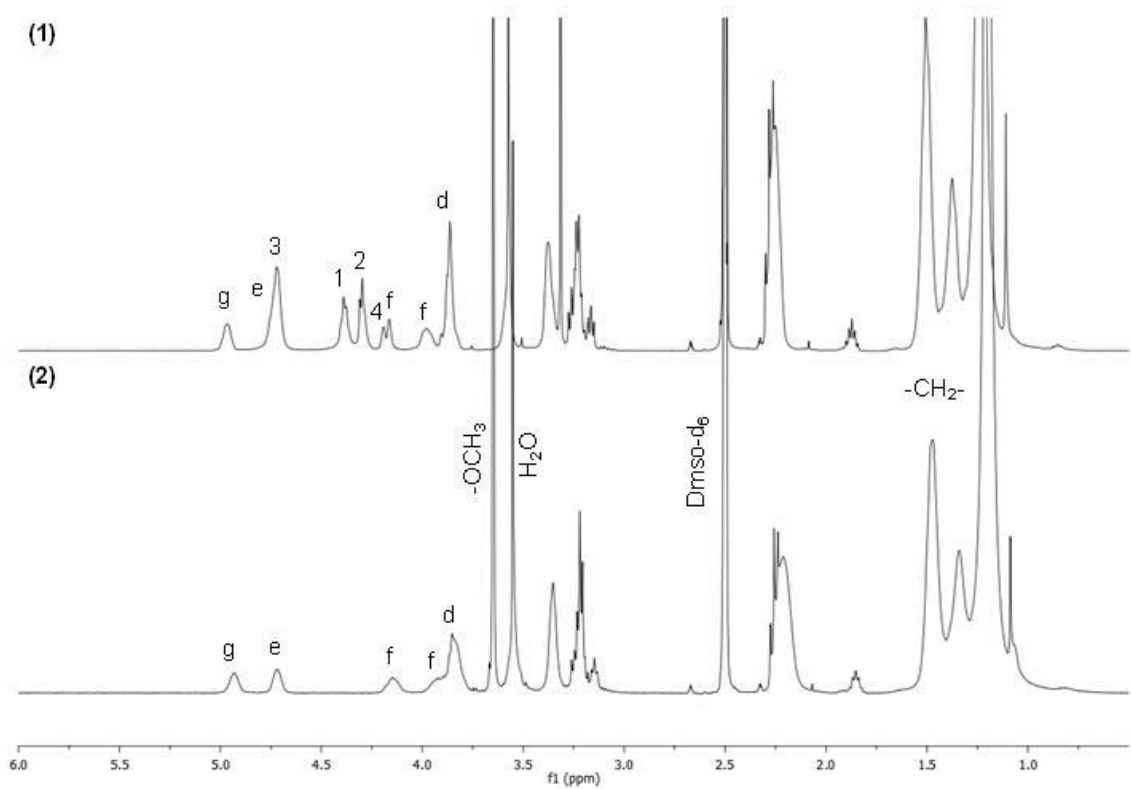

Figure SI 6. Stacked $^1$H NMR spectra of a HBPE prepared by polycondensation of M2HU in (1) DMSO-$d_6$ and in (2) DMSO-$d_6$ & a few drops of water

Table SI 1. Catalyst dependence of DB values ($DB_{Fréchet}$ and $DB_{Frey}$) for M2HS and M2HB-based HBPEs

| Entry | Monomer | Catalyst | $M_n^a$ (g.mol$^{-1}$) | $Đ^a$ | $DB_{Fréchet}^b$ | $DB_{Frey}^b$ |
|---|---|---|---|---|---|---|
| P1 | M2HS | Zn(OAc)$_2$ | 3 500 | 2.71 | 0.18 | 0.07 |
| P3 | | TBD | 4 100 | 2.46 | 0.43 | 0.35 |
| P5 | | TBD | 7 600 | >12 | 0.44 | 0.41 |
| P6 | | NaOMe | 6 100 | 3.08 | 0.32 | 0.29 |
| P7 | M2HB | Zn(OAc)$_2$ | 3 000 | 1.93 | 0.29 | 0.09 |
| P8 | | Zn(OAc)$_2$ | 5 600 | >11 | 0.26 | 0.18 |
| P9 | | TBD | 5 600 | 3.05 | 0.39 | 0.33 |
| P10 | | NaOMe | 9 200 | 3.27 | 0.30 | 0.30 |

(a) SEC in THF - calibration PS standards. (b) $^1$H NMR spectroscopy

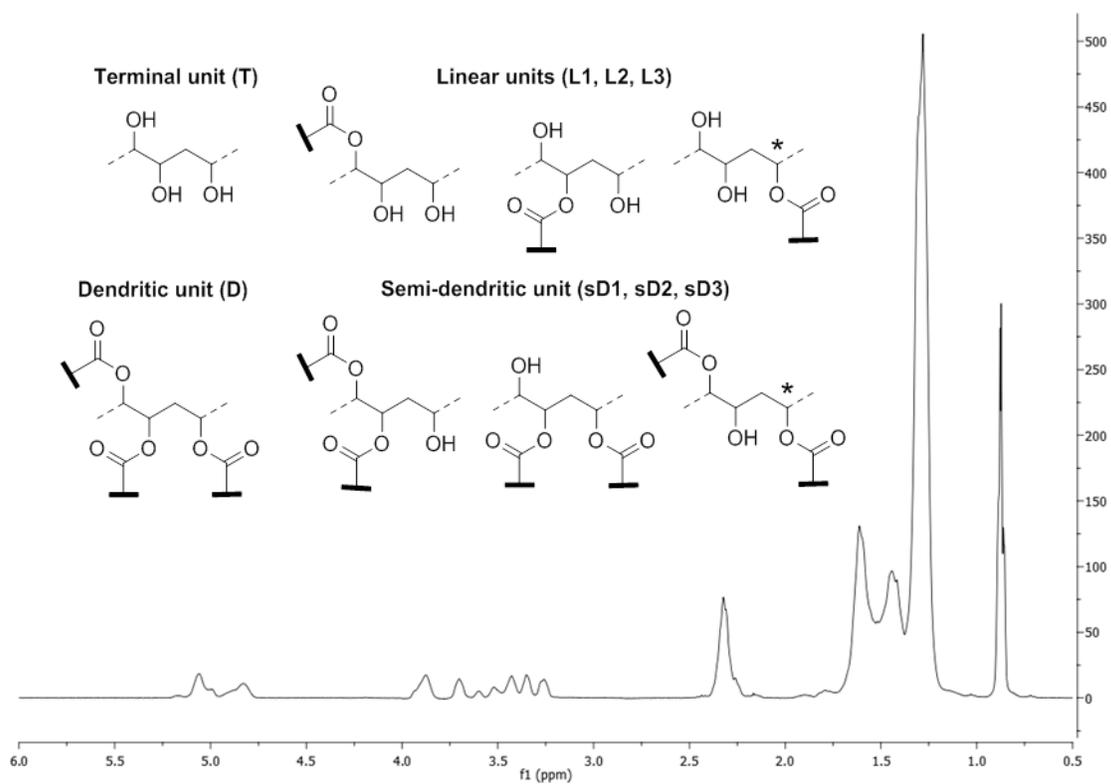

Figure SI 7. $^1$H NMR spectrum of a HBPE derived from M3HS in CDCl$_3$ and schematic representation of terminal, linear, semi-dendritic and dendritic units

Figure SI 7. Molar mass dependence of DB values for HBPEs obtained by TBD-catalyzed polycondensation of M2HS

Figure SI 8. Stacked $^1$H NMR spectra of (1) methyl 12-hydroxystearate and (2) its related linear polyester (LPE) in CDCl$_3$

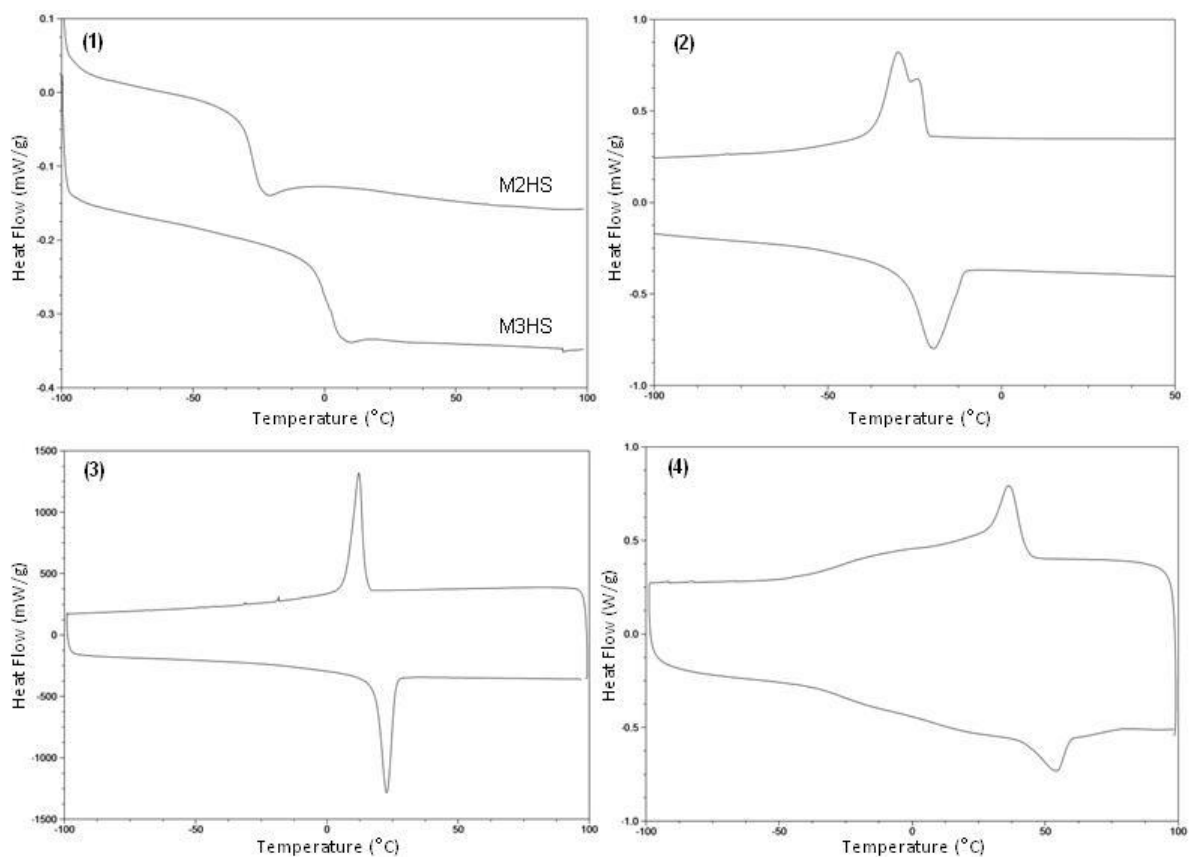

Figure SI 10. DSC thermograms at 10°C.min$^{-1}$ for HBPEs derived from (1) M2HS and M3HS, (3) M2HB and (4) M2HU, and a linear reference LPE (2)

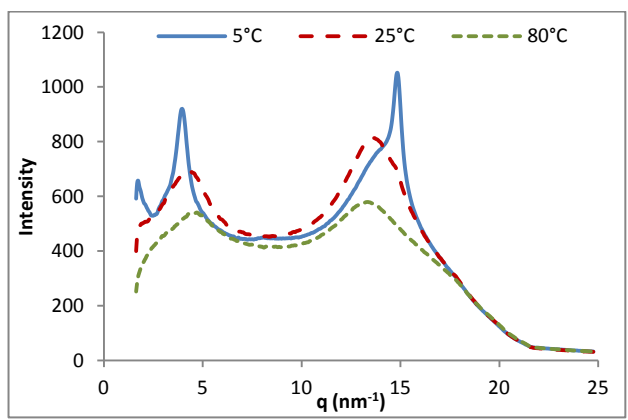

Figure SI 11. WAXS patterns at 5, 25 and 80°C of a HBPE prepared from M2HB

The semi-crystalline character of M2HB-based HBPEs was further investigated by wide-angle X rays diffraction measurements (WAXS), carried out as a function of the temperature at 5, 25 and 80°C. Figure SI 11 depicts the different WAXS patterns obtained. Results clearly confirm the semi-crystalline properties of the HBPEs. Apart from the apparent large amorphous halo, two diffraction peaks can be distinguished. Their position as well as their intensity depend on the temperature of analysis. At 80°C, *i.e.* well above the melting point, these two diffraction peaks disappeared leading to a broad amorphous halo, due to the lost of crystallinity at that temperature.

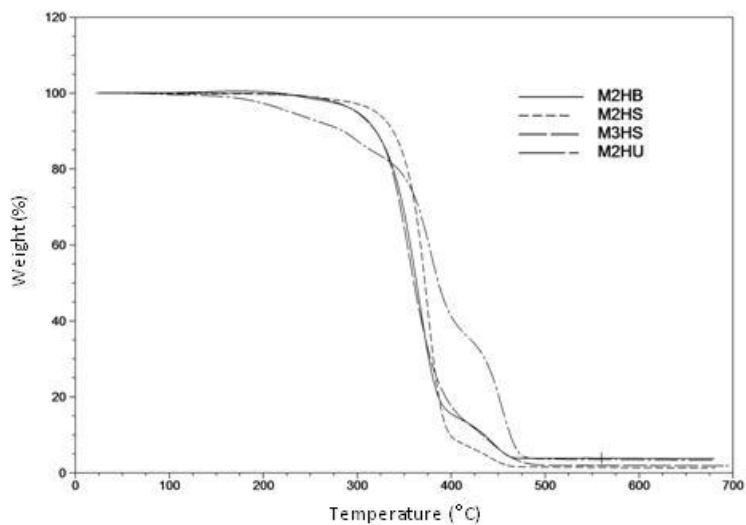

Figure SI 12. Weight loss as a function of temperature for HBPEs derived from M2HS, M2HB, M2HU and M3HS, from TGA experiments at 10°C.minutes$^{-1}$ under nitrogen atmosphere

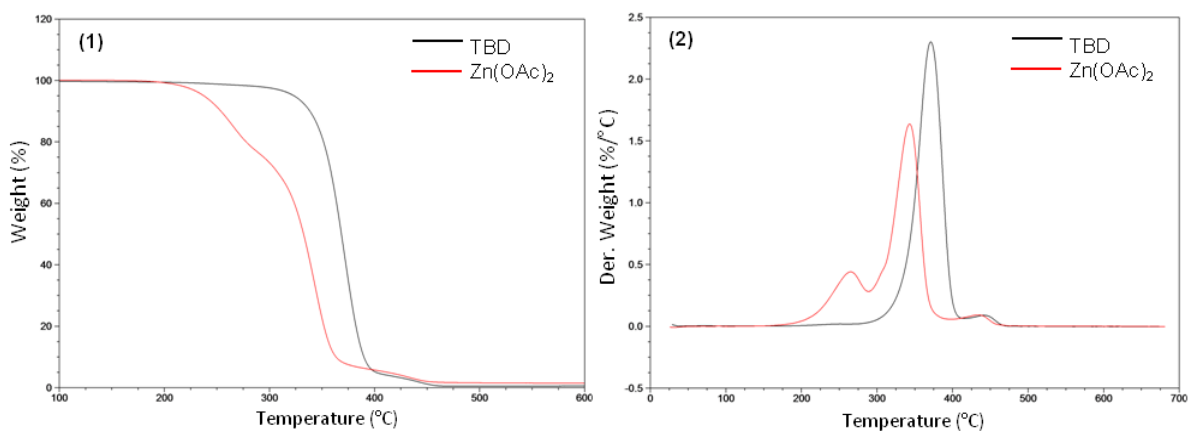

Figure SI 13. (1) Weight loss as a function of temperature (2) Derivative of weight loss with temperature for M2HS-based HBPEs using Zinc acetate and TBD as catalyst obtained from TGA experiments at 10°C/min under nitrogen atmosphere

HBPEs synthesized using $Zn(OAc)_2$ were found to display lower $T_d^{5\%}$, in between 237 and 252°C, than materials obtained with TBD (306 to 338°C). Typical decomposition profiles of M2HS-derived HBPEs prepared using both catalysts are presented in this figure. TGA derivatives of weight loss as a function of the temperature show one additional degradation step with the organometallic catalyst between 200 and 300°C. This first weight loss was assigned to the thermal zinc acetate-catalyzed depolymerization of the HBPEs knowing that all samples were characterized without further purification. This phenomenon has already been reported in the literature for poly(ethylene terephthalate) and poly(tetramethylene succinate).

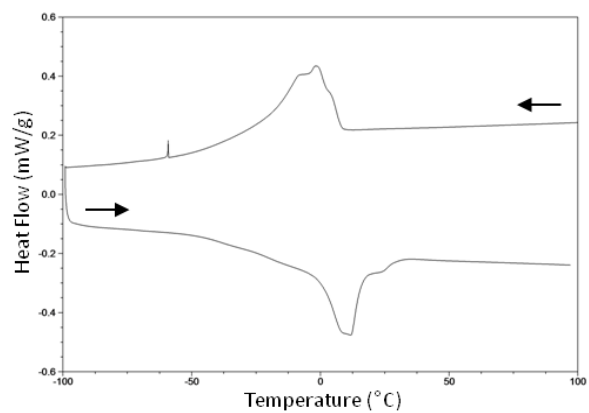

Figure SI 14. DSC trace of HBPE prepared using Zn(OAc)$_2$ after an isotherm of 15 mins at 250°C

The depolymerization phenomenon was investigated by DSC analyses. After an isotherm of 15 minutes at 250°C, the thermogram of HBPEs prepared using zinc acetate displayed an exothermic peak upon cooling at 10°C.min$^{-1}$ and one endothermic peak upon heating. Since the HBPEs were amorphous, the crystallization observed may be due to the presence of either linear segments or monomer residues within the sample. It is important to notice that this was not observed for HBPEs prepared with TBD.